\documentclass[epj]{svjour}
\usepackage{graphics}
\begin{document}
\title{Universal Centrality and Collision Energy Trends for $v_2$ Measurements From 2D Angular Correlations}
\author{David Kettler\inst{1} on behalf of the STAR collaboration
}                     
%
%
\titlerunning{Universal Centrality and Collision Energy Trends}
\institute{University of Washington}
\date{Received: date / Revised version: date}
%
\abstract{
We have measured the $p_t$-integrated quadrupole component of two-particle azimuth correlations (related to quantity $v_2$, denoted in this case by $v_2 \{ 2D\}$) via two-dimensional (2D) angular autocorrelations on $(\eta , \phi )$ for unidentified hadrons in Au-Au collisions at 62 and 200 GeV.  The 2D autocorrelation provides a method to remove non-quadrupole contributions to $v_2$ (conventionally termed ``nonflow'') under the assumption that such processes produce significant dependence on pair-wise relative $\eta$ within the detector acceptance.  We hypothesize, based on empirical observations, that non-quadrupole contributions are dominated by minijets or minimum-bias jets.  Using the optical Glauber eccentricity model for initial-state geometry we find simple and accurate universal energy and centrality trends for the quadrupole component. Centrality trends are determined only by the initial state (impact parameter $b$ and center-of-mass energy $\sqrt{s_{NN}}$ ).
There is no apparent dependence on evolving system dynamics (e.g., equation of state or number of secondary collisions).
Our measurements of the quadrupole and non-quadrupole components have implications for the contributions to $v_2$.  They suggest that the main source of the difference between $v_2 \{ 2\}$ and $v_2 \{ 4\}$ (or $v_2 \{ 2D\}$) is measured properties of minijets.
\PACS{
      {PACS-key}{discribing text of that key}   \and
      {PACS-key}{discribing text of that key}
     } 
} 
\maketitle
\section{Introduction}
\label{intro}

The azimuthal quadrupole in the distribution of particles in heavy-ion collisions, or ``elliptic flow'' has long been described in terms of hydrodynamical phenomena \cite{classicv2}.  Interpreting traditional elliptic flow measurements is difficult because of issues with ``nonflow'', particle correlations caused by phenomena such as minijets, resonances, and HBT.  We present a two-dimensional autocorrelation analysis in which the quadrupole (related to elliptic flow) and non-quadrupole (caused by non-flow) components of two-particle correlations are measured simultaneously.

In this analysis we use the structure of nonflow correlations on pair-wise relative pseudorapidity to distinguish \emph{geometrically} the azimuth quadrupole moment, attributed to elliptic flow, from nonflow effects, which we believe to be due primarily to minijet correlations---fragments from low-$Q^2$ partons, mainly gluons \cite{conf2005,mj1,mj2,mj3,frag}.  Quadrupole amplitudes are obtained for the first time from fits to 2D angular autocorrelations on azimuth and pseudorapidity.  Measurements of the quadrupole moment over a broad range of centralities and energies provide qualitatively new insights into the phenomenon of elliptic flow.  Quadrupole amplitudes follow simple systematic trends on centrality and energy described by only two initial-state parameters for all systems down to 13 GeV.

\section{Azimuth Autocorrelations}
\label{ewac}
We construct minimum-bias angular autocorrelations by considering all possible pairs of particles in an event.  In general one can consider structures in the pair density on 6D momentum space $(p_{t1} , \eta_1 , \phi_1 , p_{t2} , \eta_2 , \phi_2 )$.  In this analysis we study $p_t$-integrated correlations on the angular subspace $( \eta_1 , \phi_1 , \eta_2 , \phi_2 )$, where the angle parameters for relativistic collisions are pseudorapidity $\eta$ (related to polar angle $\theta$) and azimuth $\phi$.  In this section we will concentrate on autocorrelations on only azimuth.

For the event-wise azimuth density $\rho ( \phi )$ the event-wise two-particle density is defined as $\rho _2 ( \phi _1 , \phi _2 ) = \rho ( \phi _1 ) \rho ( \phi _2 )$.  The azimuth autocorrelation is then just the projection by averaging of the two-particle density onto the difference axis $\phi _\Delta \equiv \phi _1 - \phi _2$ \cite{fluctinv}.
Consider a binned system with bin width $\delta$.  The autocorrelation is then
\begin{eqnarray}
r_{\rho}(\phi_{\Delta}) &=& \frac{1}{2\pi \delta} \int_{-\pi}^{\pi} d\phi_{\Sigma}^{\prime} \int_{\phi_\Delta - \delta /2}^{\phi_\Delta + \delta /2} d\phi_{\Delta}^{\prime} \, \rho (\phi_{1}^{\prime}) \rho (\phi_{2}^{\prime}) \\
&=& \frac{1}{\pi \delta} \int_{\phi_\Delta - \delta /2}^{\phi_\Delta + \delta /2} d\phi_{\Delta}^{\prime} \int_{\frac{-\pi + \phi_{\Delta}^{\prime}}{2}}^{\frac{\pi + \phi_{\Delta}^{\prime}}{2}} d\phi_{1}^{\prime} \, \rho (\phi_{1}^{\prime}) \rho (\phi_{1}^{\prime}-\phi_{\Delta}^{\prime}) \nonumber
.
\end{eqnarray}
If we use the periodicity of $\phi$ then in the limit where the bin width goes to zero we get
\begin{eqnarray}
r_{\rho}(\phi_{\Delta}) &=& \frac{1}{2\pi} \int_{-\pi}^{\pi} \rho (\phi^{\prime} ) \rho (\phi^{\prime} - \phi_{\Delta}) d \phi^{\prime} ,
\end{eqnarray}

which is the standard definition of the autocorrelation.

While the autocorrelation procedure is mathematically well-defined for a single event, in practice the number of particles in an event is far too small---even in central Au-Au collisions---to measure two-particle densities with any statistical accuracy.
It is therefore essential to construct an average of the autocorrelations over many events.
The stationarity property of two-particle correlation structures \cite{fluctinv} ensures that even in an ensemble average the information on the difference axis of a two-particle density---the autocorrelation---is not lost.
The key feature of using autocorrelations is that all of the structure in an event is preserved, which includes both flow and nonflow effects.  We will discuss how to seperate the two in the next section.

Contrast this with the standard (event plane) technique of measuring $v_2$ \cite{poskvol} which depends on shifting the single-particle densities according to the estimate of the reaction plane and averaging the shifted single-particle densities.  If the reaction plane is known perfectly then this is an effective approach as the nonflow should not be correlated with the reaction plane and when averaged over many events will contribute as a flat background.  However, when the reaction plane is estimated using particles in the event then there are correlations between the particles used to estimate the reaction plane and the particles used to measure the azimuthal anisotropy making the event-plane technique effectively another type of two-particle correlation \cite{azstruct}.  This is the origin of nonflow contributions in the event-plane method.

A proper statistical measure helps us understand how these correlations scale.  We will refer to the pair density as $\rho$ and define $\rho_{\textrm{ref}}$ to be a reference pair density constructed from mixed events.  We can construct a \emph{per-pair density ratio} $\Delta \rho / \rho_{\textrm{ref}} = \rho / \rho_{\textrm{ref}} - 1$ which varies as $1/n_{ch}$ with changing system size in the absence of other physical changes.  We prefer a statistical measure whose variation reflects deviations from the null hypothesis based on N-N superposition, the \emph{per-particle density ratio} $\Delta \rho / \sqrt{\rho_{\textrm{ref}}}$ (Pearson's normalized covariance coverted to a density ratio) which exhibits the desired properties, since $\sqrt{\rho_{\textrm{ref}}} \propto n_{ch}$.

\section{2D Autocorrelations}
\label{ddac}
\begin{figure*}
\resizebox{1.00\textwidth}{!}{%
  \begin{tabular}{cccc}
  \includegraphics{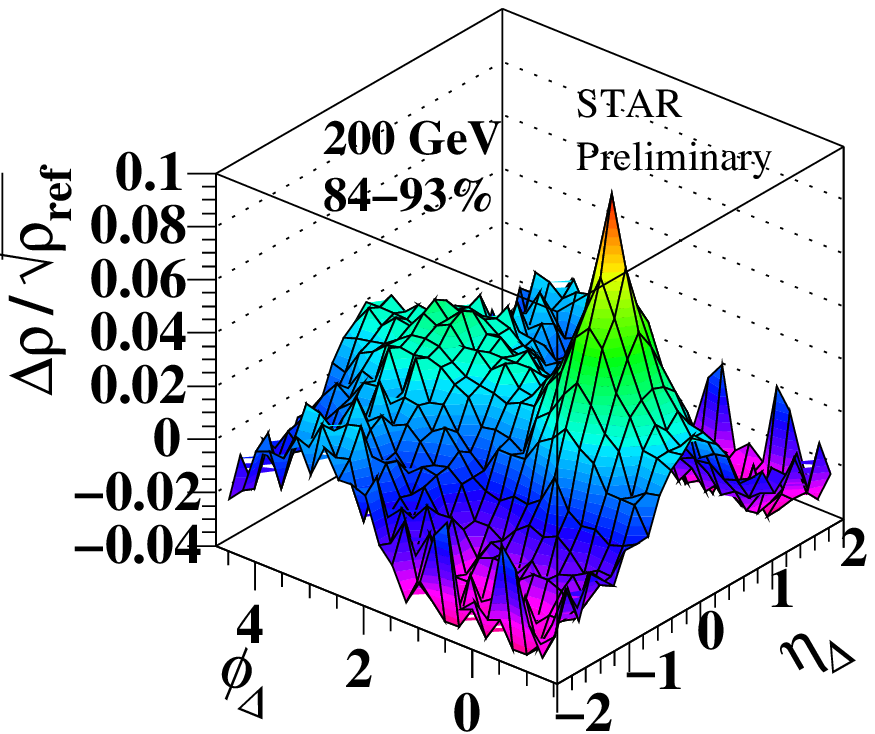} &
  \includegraphics{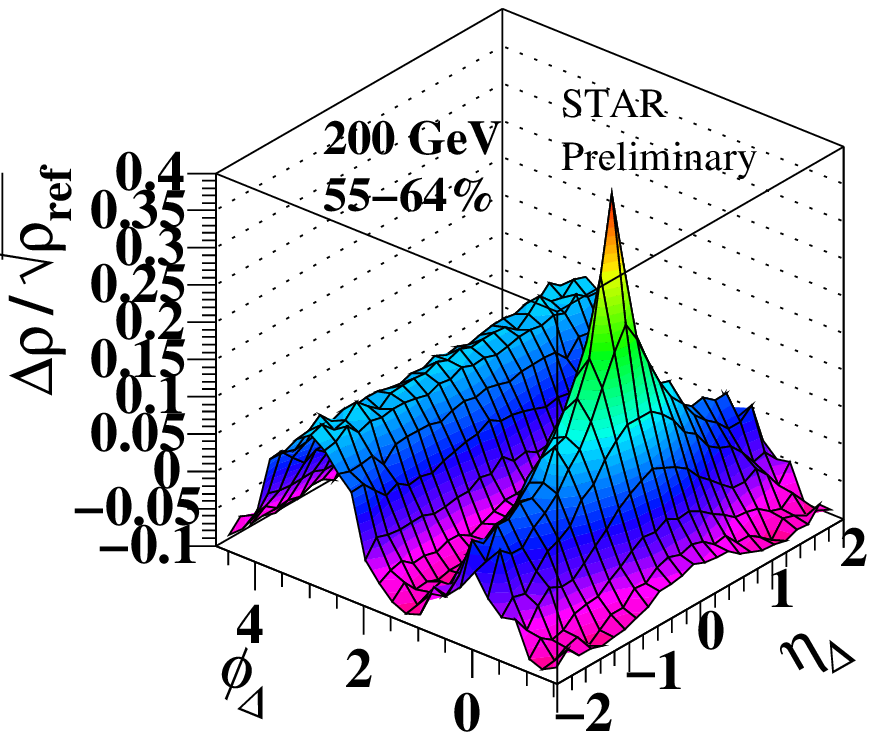} &
  \includegraphics{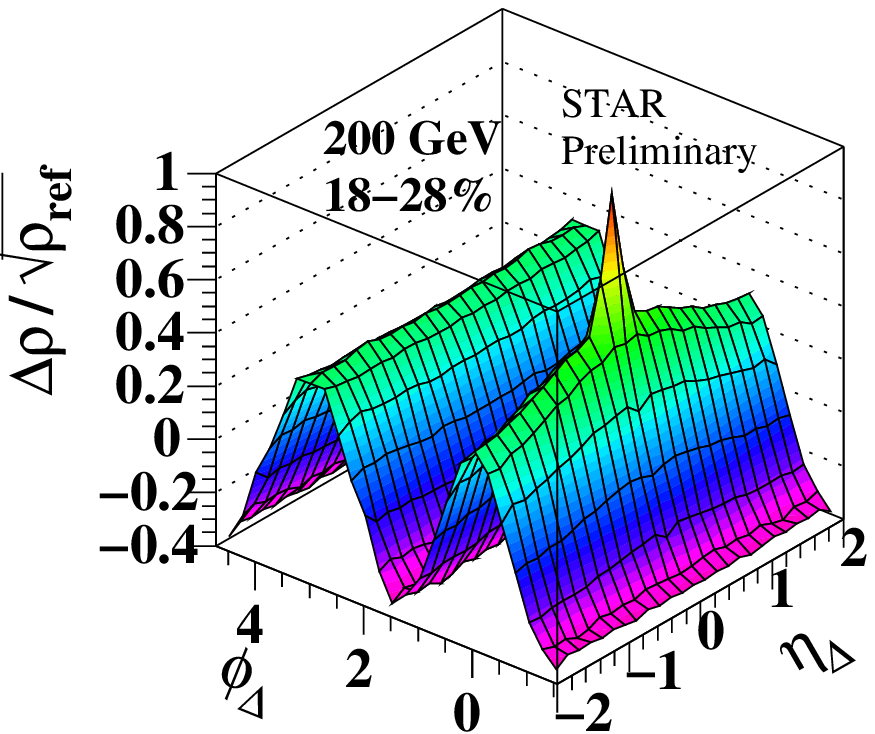} &
  \includegraphics{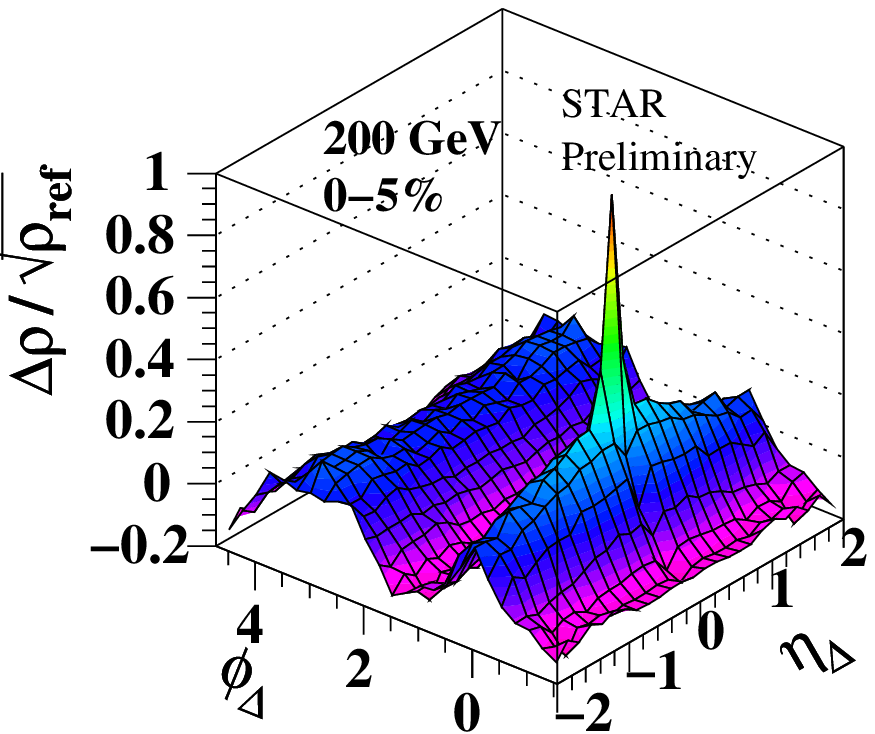} \\
  \includegraphics{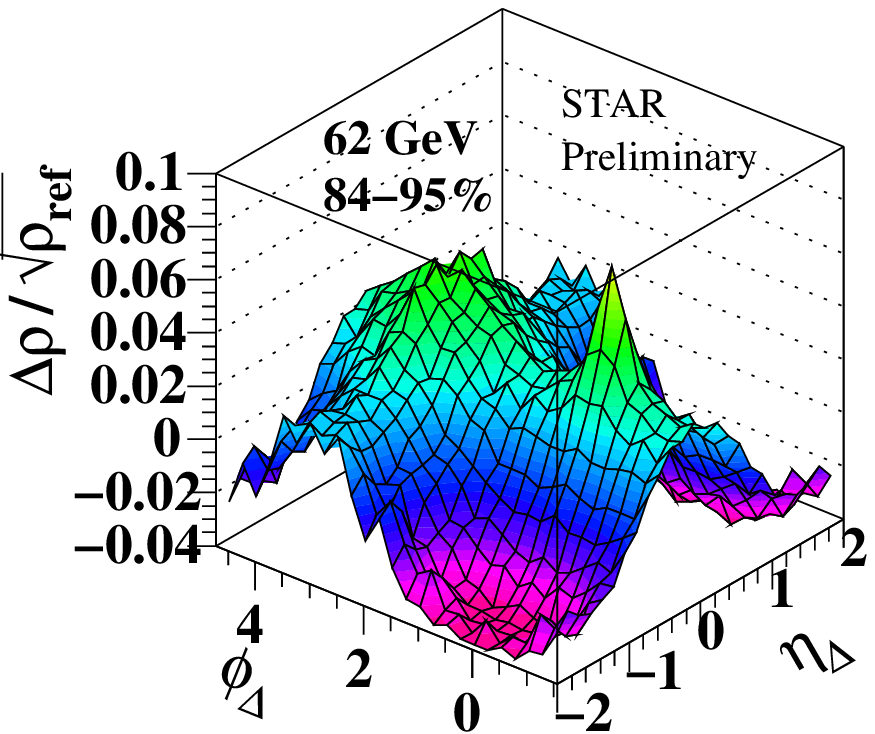} &
  \includegraphics{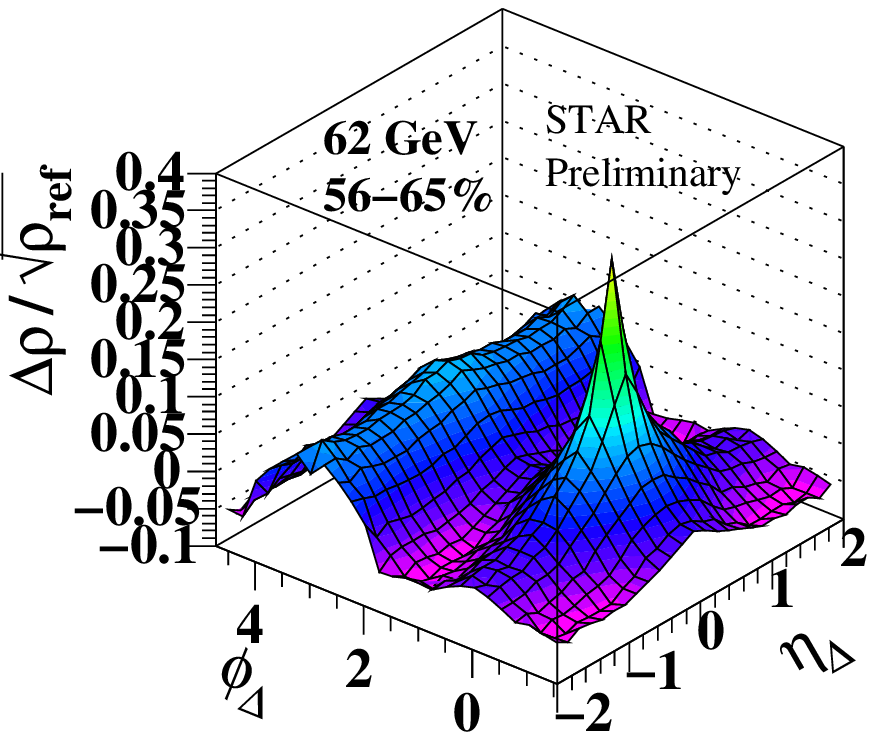} &
  \includegraphics{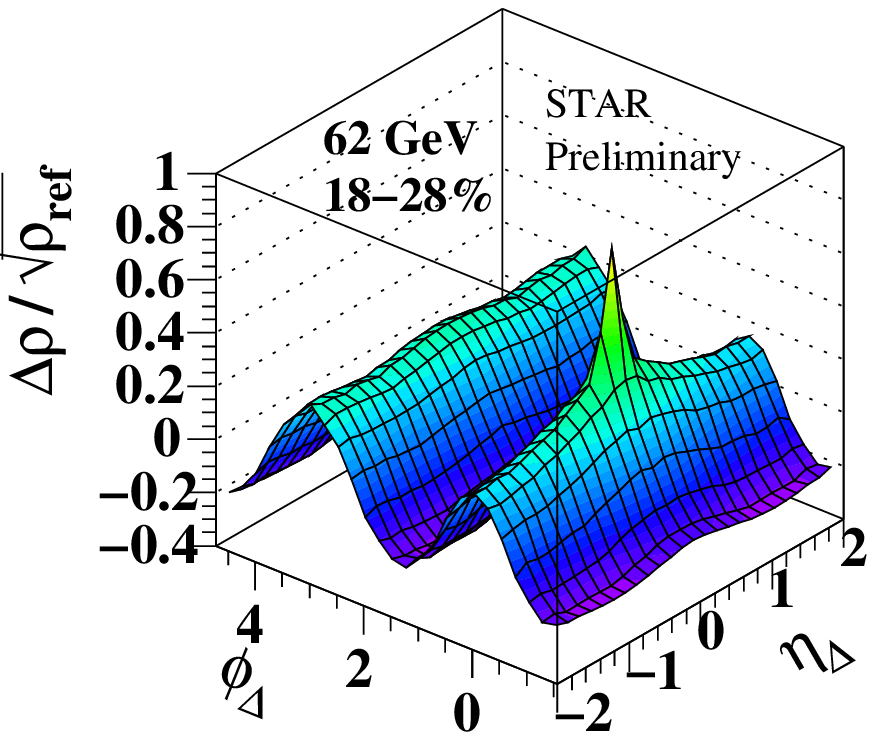} &
  \includegraphics{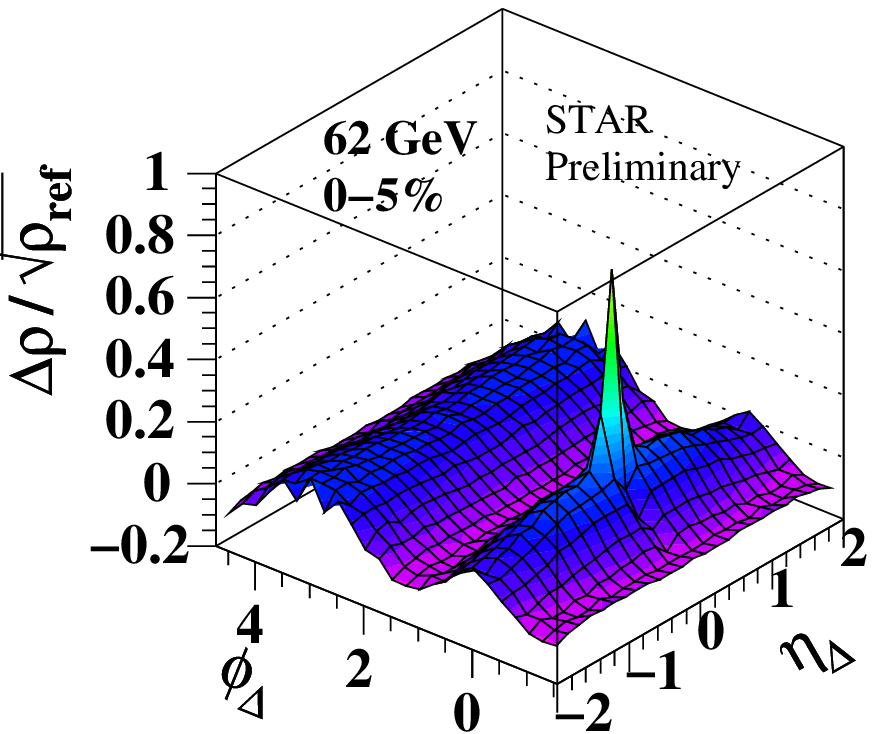}
  \end{tabular}
}
\caption{Perspective views of two-dimensional charge-independent correlations $\Delta \rho / \sqrt{\rho_{\textrm{ref}}}$ on $(\eta_\Delta , \phi_\Delta )$ for Au-Au collisions at $\sqrt{s_{NN}}=200$ and 62 GeV (upper and lower rows respectively).  Centrality increases left-to-right from most-peripheral to most-central.  Corrected total cross-section fractions are noted for each centrality bin \cite{centralities}.}
\label{fig:corrs}       
\end{figure*}

We now consider the more general problem of angular correlations on $(\eta_1, \eta_2, \phi_1, \phi_2)$.  By following the procedure described in the previous section simultaneously on $\phi$ and $\eta$ we can construct 2D angular autocorrelations on $(\eta_\Delta , \phi_\Delta )$.  Examples from 62 and 200 GeV minimum-bias Au-Au collisions are shown in Fig.~\ref{fig:corrs}.

$p_t$-integrated 2D angular autocorrelations contain two types of structure: $\eta_\Delta$-dependent peaks and $\eta_\Delta$-indepen\-dent sinusoids $\cos (\phi_\Delta )$ and $\cos (2 \phi_\Delta )$, where the $\cos (2 \phi_\Delta )$ sinusoid is related to elliptic flow, and the $m=2$ Fourier component of the remaining structure is the source of nonflow in a standard two-particle analysis.  We therefore decompose the density ratio into a part varying with $\eta_\Delta$ (nonflow, identified with subscript nf) and a part independent of $\eta_\Delta$ composed of the $m=1 , 2$ terms in a Fourier series
\begin{eqnarray}
\frac{\Delta \rho}{\sqrt{\rho_{\textrm{ref}}}} \equiv \frac{\Delta \rho_{\textrm{nf}}}{\sqrt{\rho_{\textrm{ref}}}} (\eta_\Delta , \phi_\Delta ) + 2 \sum _{m=1} ^2 \frac{\Delta \rho [m]}{\sqrt{\rho_{\textrm{ref}}}} \cos (m \phi_\Delta )
\end{eqnarray}

The correlation data for each energy and centrality were fitted with a six-component model function motivated by structures observed in data from 200 GeV p-p collisions \cite{conf2005}.  The only addition necessary to describe heavy ion data is the quadrupole component.

The model function includes: a same-side 2D Gaussian on $(\eta_\Delta , \phi_\Delta )$, and $\eta_\Delta$-independent negative dipole $-\cos (\phi_\Delta )$, an $\eta_\Delta$-independent quadrupole $\cos (2 \phi_\Delta )$ (identified with elliptic flow \cite{azstruct}), a $\phi_\Delta$-independent 1D Gaussian on $\eta_\Delta$ (related to participant-nucleon fragmentation), a narrow same-side 2D exponential on $(\eta_\Delta , \phi_\Delta )$ (modeling quantum correlations and conversion electron pairs), and a constant normalization offset.  The model function is expressed as
\begin{eqnarray}
F &=& A_{\phi_\Delta} \cos (\phi_\Delta ) + A_{2 \phi_\Delta} \cos (2\phi_\Delta ) + A_0 e^{- \frac{1}{2} \left( \frac{\eta_\Delta}{\sigma_0} \right) ^2} \nonumber \\
&+& A_1 e^{- \frac{1}{2} \left\{ \left( \frac{\phi_\Delta}{\sigma_{\phi_\Delta}} \right) ^2 + \left( \frac{\eta_\Delta}{\sigma_{\eta_\Delta}} \right) ^2 \right\} } \nonumber \\
&+& A_2 e^{- \left\{ \left( \frac{\phi_\Delta}{w_{\phi_\Delta}} \right) ^2 + \left( \frac{\eta_\Delta}{w_{\eta_\Delta}} \right) ^2 \right\} ^{1/2} } + A_3 .
\label{eqn:fitfunc}
\end{eqnarray}

\section{Quadrupole Centrality and Energy Systematics}
\label{system}

This analysis is based on 6.7M and 1.2M Au-Au collisions at $\sqrt{s_{NN}}=62.4$ and 200 GeV respectively, observed with the STAR time projection chamber \cite{tpcs}. The acceptance was defined by transverse momentum $p_t > 0.15 \textrm{ GeV}/c$, $| \eta | < 1$ and $2\pi$ azimuth. Au-Au collision centrality was defined as in \cite{centralities}. Minimum-bias event samples were divided into 11 centrality bins: nine $\sim 10\%$ bins from 100\% to 10\%, the last 10\% divided into two $\sim 5\%$ bins.  The corrected centrality of each bin as modified by tracking and event vertex inefficiences was determined by relating the mean $n_{ch}^{1/4}$ of the bin to path length $\nu$ according to running-integral procedures in \cite{centralities}.

The quadrupole term of our fits can be related to the conventional $v_2$ measure according to \cite{azstruct}
\begin{eqnarray}
\frac{\Delta \rho [2]}{\sqrt{\rho_{\textrm{ref}}}} \equiv \frac{\bar{n} v_2 ^2 \{ 2D\} }{2 \pi}.
\end{eqnarray}

\begin{figure}
\resizebox{0.50\textwidth}{!}{%
  \includegraphics{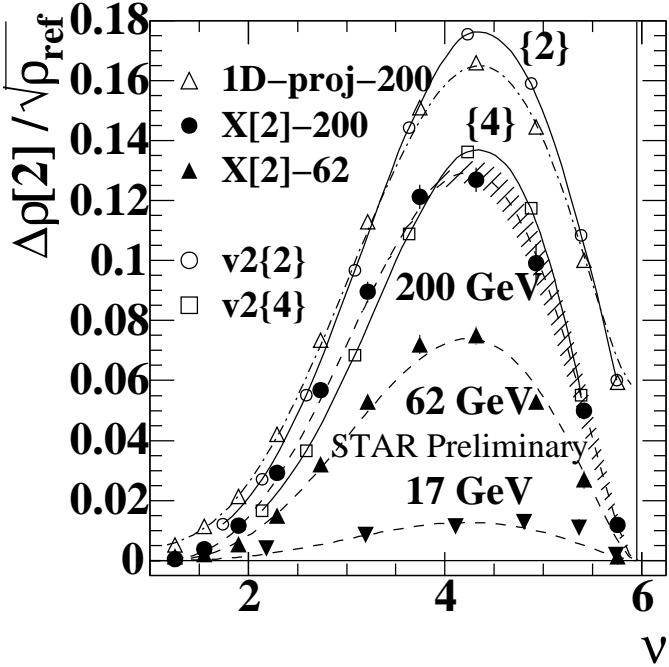}
}
\caption{Quadrupole component $\Delta \rho [2] / \sqrt{\rho_{\textrm{ref}}} \equiv X[2]$ of the autocorrelation density ratio inferred from model fits to 2D angular autocorrelations as in Fig.~\ref{fig:corrs} (solid dots and upright triangles) \emph{vs.} centrality $\nu$ (mean participant path length).}
\label{fig:centralitya}       
\end{figure}

\begin{figure}
\resizebox{0.50\textwidth}{!}{%
  \includegraphics{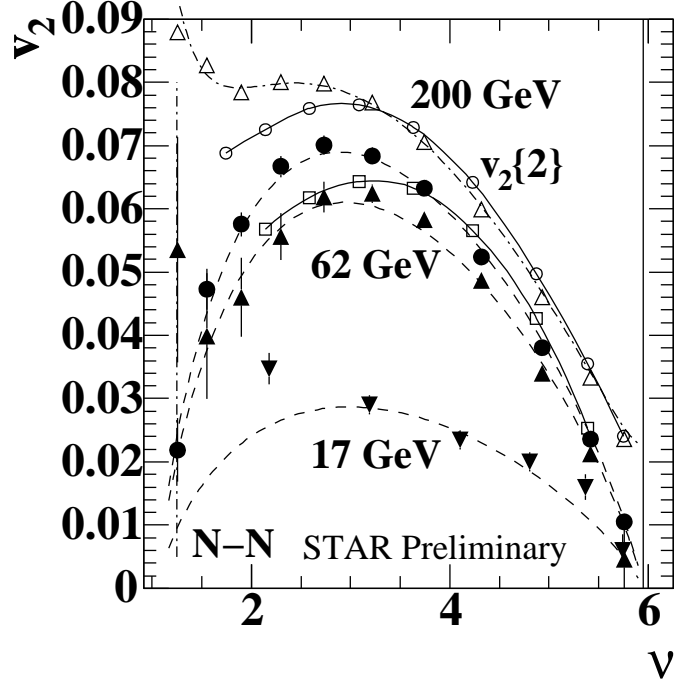}
}
\caption{$v_2 \{ 2D\}$ from 2D autocorrelations (solid points). $v_2$ data at $\sqrt{s_{NN}} = 17$ GeV are inverted solid triangles in both panels. The open symbols are explained in the text. Dashed curves have the same form for all energies above 13 GeV (see text). Dash-dot curves include ``nonflow.'' The vertical solid lines at $\nu \sim 5.95$ indicate $\nu$ for $b=0$. The dash-dot line at $\nu = 1.25$ corresponds to N-N collisions \cite{centralities}.}
\label{fig:centralityb}       
\end{figure}

Fig.~\ref{fig:centralitya} summarizes Eq.~(\ref{eqn:fitfunc}) 2D fit results for $\Delta \rho [2] / \sqrt{\rho_{\textrm{ref}}}$, with corresponding values of $v_2 \{ 2D\}$ in Fig.~\ref{fig:centralityb} for comparison to previous analyses.  The left panel shows fit results for 200 GeV (solid dots) and 62 GeV (solid upright triangles) data: strong increase with centrality to mid-central collisions followed by reduction to zero for central collisions.
$v_2 \{ \textrm{EP}\}$ data from NA49 (inverted solid triangles) \cite{na49} provide a reference for energy-dependence systematics.

Published data for two-particle $v_2 \{ 2\}$ (open circles) and four-particle cumulant $v_2 \{ 4\}$ (open squares) at 200 GeV \cite{v24} are compared to $v_2 \{ 2D\}$ (solid points) from this 2D autocorrelation analysis.  The $v_2 \{ 1D\}$ (open triangles) are fits of $\cos (2 \phi_\Delta )$ to 1D projections onto $\phi_\Delta$ of the 200 GeV 2D autocorrelations, roughly consistent with the $v_2 \{ 2\}$ analysis as expected \cite{azstruct,quadspec} and substantially larger than the 2D fits.  That nonflow offset is expected in a conventional 1D flow analysis: the difference between open triangles and solid dots is predominantly the $m=2$ Fourier component of the same-side minijet peak \cite{quadspec}.  $v_2 \{ 4\}$ is expected to eliminate nonflow assuming that elliptic flow is a collective property of many particles, whereas ``nonflow'' describes independent ``clusters'' of a few particles \cite{clusters}.  The open squares in Fig.~\ref{fig:centralitya} are closer to the 2D analysis, but systematic deviations outside published uncertainties remain.  The difference $v_2 \{ 2\} - v_2 \{ 4\}$ has been explained by a combination of nonflow and flow fluctuation effects \cite{v2fluct}.  However, we see a similar difference in $v_2 \{ 1D\} - v_2 \{ 2D\}$, which are both two-particle correlations and respond identically to flow fluctuations.  We then conclude that the main source of the differences in these quantities is the same-side minijet peak, not $v_2$ fluctuations.

The comparison of $v_2 \{ 2D\}$ and $v_2 \{ 1D\}$ allows us to calculate the contribution of nonflow to $v_2 \{ 1D\}$
in 200 GeV Au-Au collisions to be between 20-100\%, depending on centrality.  The problem is largest in the most central collisions, since not only is $v_2 \{ 2D\}$ going to zero but the amplitude of the nonflow term is at a maximum.

\section{Eccentricity}
\label{scaling}

Interpretations of the Au-Au azimuth quadrupole require understanding the geometry of the initial collision system as characterized by the eccentricity \cite{ecc} defined by
\begin{eqnarray}
\epsilon = \frac{\left< y \right>^{2} - \left< x \right>^{2}}{\left< y \right>^{2} + \left< x \right>^{2}},
\end{eqnarray}
where $x$ and $y$ are coordinates in the plane perpendicular to the beam axis and $x$ refers to the direction along the reaction plane.
However there are major uncertainties about the most appropriate way to calculate eccentricities for nuclear collisions.

In the optical model the eccentricity is approximated by a continuous transverse density profile, typically calculated from a Woods-Saxon distribution.  We shall refer to this as $\epsilon_{\textrm{opt}}$.
The participant model of eccentricity uses a Monte Carlo Glauber calculation that builds nuclei by randomly placing nucleons according to a Woods-Saxon distribution \cite{part}.  We shall refer to this as $\epsilon_{\textrm{part}}$.  This model is favored by many both because of the apparent scaling the quantity $v_2 / \epsilon_{\textrm{part}}$ exhibits between Cu-Cu and Au-Au collision systems \cite{phobscale} and because it provides a mechanism for flow fluctuations.  However, in \cite{phobscale} $v_2$ was measured with the event plane method which does not adequately account for nonflow effects which are large in central collisions and as stated in section~\ref{system} there is reason to believe that flow fluctuations may not play a large role.
If nonflow contributions have not been completely removed then the scaling behavior can change when they are.

\begin{figure}
\resizebox{0.50\textwidth}{!}{%
  \includegraphics{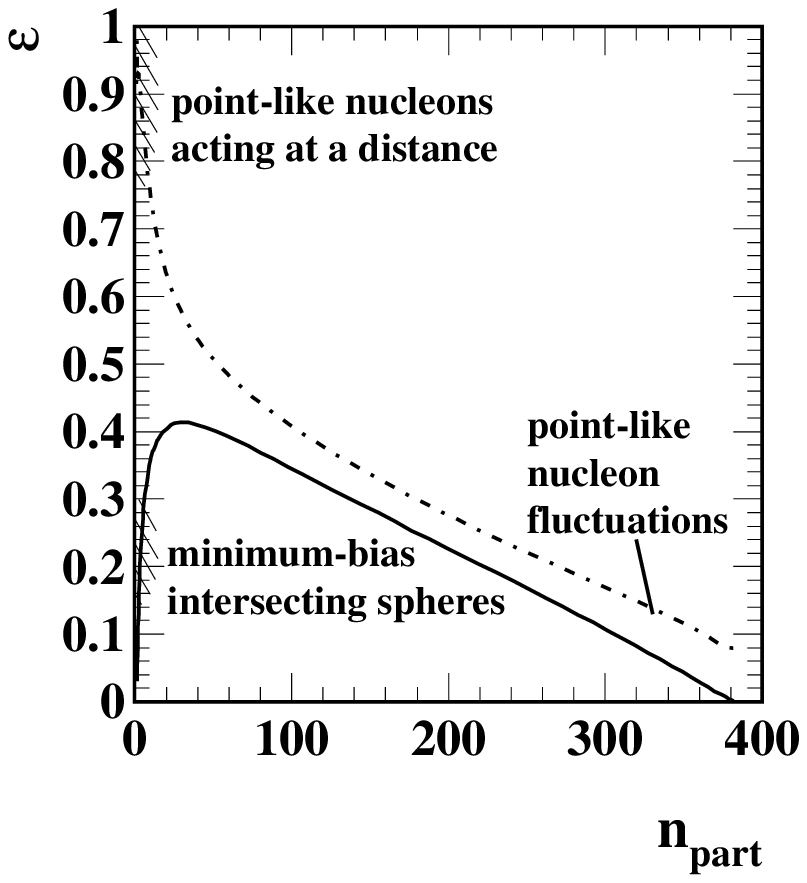}
  \includegraphics{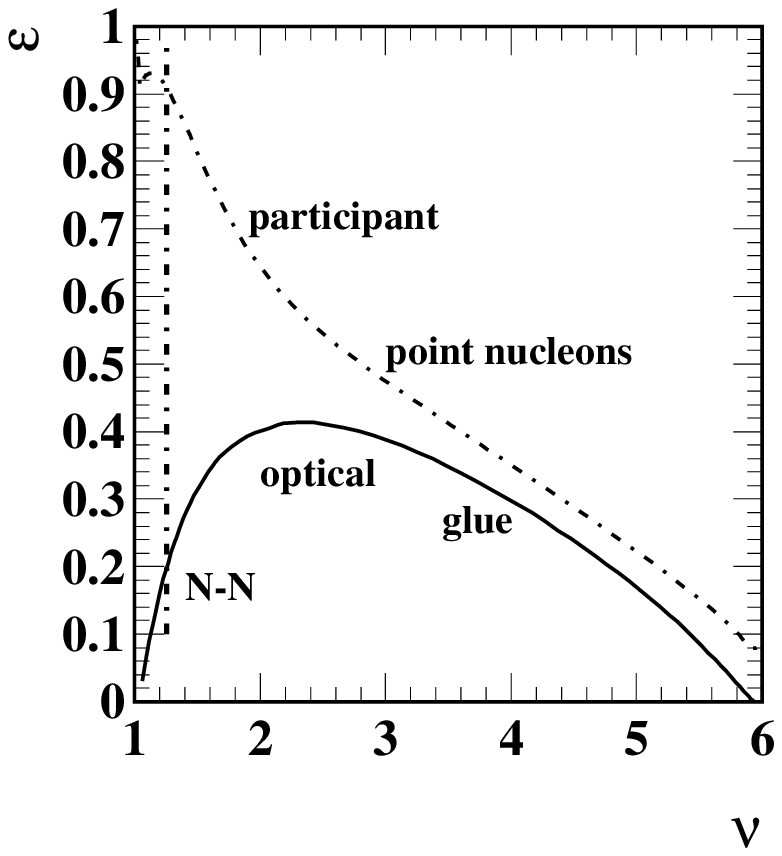}
}
\caption{Left Panel: Optical eccentricity $\epsilon_{\textrm{opt}}$ (solid curve) and participant eccentricity $\epsilon_{\textrm{part}}$ (dash-dot curve) \cite{quadspec} \emph{vs.} participant nucleon number $n_{\textrm{part}}$.  Right Panel: The same curves as Fig.~\ref{fig:ecca} \emph{vs.} mean participant path length $\nu$.  The vertical dash-dot line denotes the mean value of $\nu$ for N-N collisions \cite{centralities}}
\label{fig:ecca}       
\end{figure}

Figure~\ref{fig:ecca} shows the differences between the participant and optical eccentricities.
The optical calculation was done according to the Jacobs and Cooper Wounded nucleon method with a Woods-Saxon potential \cite{jaccoop}, however the calculation was done using a nucleon-nucleon cross section of 42 mb which is more appropriate for RHIC collision energies.  We found that there is a simple parametrization of this calculation given by
\begin{eqnarray}
\epsilon_{\textrm{opt}} (n_{\textrm{bin}}) = \frac{1}{5.68} \log_{10} \left( \frac{3 n_{\textrm{bin}}}{2} \right) ^{0.96} \log_{10} \left( \frac{1136}{n_{\textrm{bin}}} \right) ^{0.81}
\end{eqnarray}

The large value of $\epsilon_{\textrm{part}}$ for the most peripheral collisions is unphysical as it implies that the nucleons are point-like objects acting at a distance.  The nonzero $\epsilon_{\textrm{part}}$ for the most central collisions seems to give desirable scaling properties when compared to event-plane $v_2$ \cite{phobscale}.  However, our measurement of $v_2 \{ 2D\}$ goes to zero for central collisions, which favors $\epsilon_{\textrm{opt}}$.  Given the possibility that elliptic flow is a ``long-wavelength'' probe that is dependent only on initial-state collision parameters \cite{quadspec}, we will use $\epsilon_{\textrm{opt}}$.

\section{Quadrupole Scaling}
\label{trends}

Returning to the quadrupole data in Fig.~\ref{fig:centralitya} we note two interesting features: 1) all energies are described by the same centrality variation (dashed curves), and 2) the energy dependence of the quadrupole amplitude is propotional to $\log (\sqrt{s_{NN}}/13 \textrm{ GeV})$.  A similar energy dependence was observed for $\left< p_t \right>$ fluctuations/correlations attributed to minijets \cite{mj3}.

\begin{figure}
\resizebox{0.50\textwidth}{!}{%
  \includegraphics{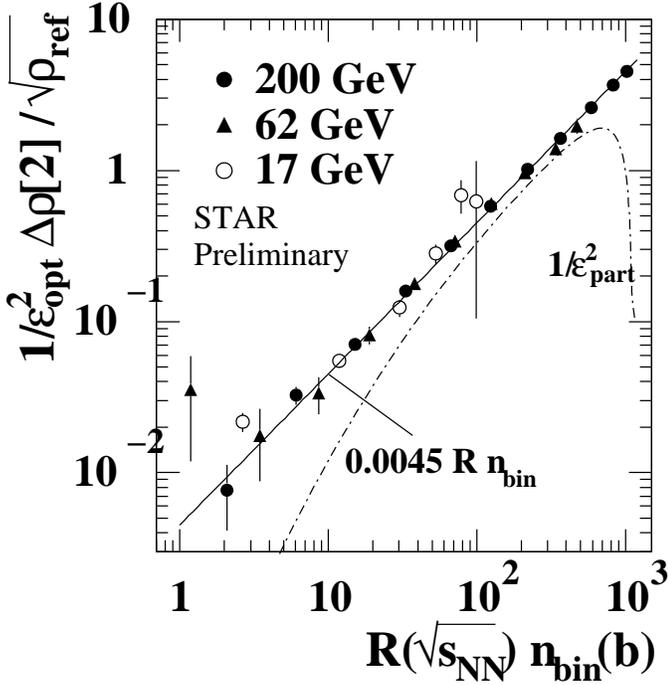}
}
\caption{The azimuth quadrupole component divided by \emph{optical} $\epsilon^2$ \emph{vs.} energy-dependent factor $R(\sqrt{s_{NN}})$ times 200 GeV Au-Au binary-collision number $n_{\textrm{bin}}(b)$. The data are consistent with simple proportionality over three decades.  The dash-dot curve is obtained by using $\epsilon_{\textrm{part}}$.}
\label{fig:energya}       
\end{figure}

In Fig.~\ref{fig:energya} $1 / \epsilon_{\textrm{opt}}^2 \cdot \Delta \rho [2] / \sqrt{\rho_{\textrm{ref}}}$ is plotted \emph{vs.} the number of binary collisions, $n_{\textrm{bin}}$, times a parameter describing the energy scaling
\begin{eqnarray}
R(\sqrt{s_{NN}}) \equiv \log \{ \sqrt{s_{NN}} / 13 \textrm{ GeV} \} / \log (200/13).
\end{eqnarray}
For all Au-Au collisions the data can be described by
\begin{eqnarray}
\frac{\Delta \rho [2]}{\sqrt{\rho_{\textrm{ref}}}} (n_{\textrm{bin}} , \sqrt{s_{NN}}) = A R(\sqrt{s_{NN}}) n_{\textrm{bin}} \epsilon_{\textrm{opt}}^2 (n_{\textrm{bin}})
\label{eqn:scaling}
\end{eqnarray}
where coefficient $A$ is defined by $1000A = 4.5 \pm 0.2$. Deviations of 17 GeV event-plane $v_2$ from the linear trend are consistent with expected contributions from minijets \cite{quadspec}.  Equation~(\ref{eqn:scaling}) accurately describes measured $p_t$-integrated azimuth quadrupole moments in heavy ion collisions for all centralities down to N-N collisions and all energies down to $\sqrt{s_{NN}} \sim 13 \textrm{ GeV}$. Transformed to each plotting space it defines the dashed curves in Fig.~\ref{fig:centralitya}. The dash-dot curves passing through the 200 GeV 1D projection points in Fig.~\ref{fig:centralitya} are obtained by adding ``nonflow'' parametrization $g_2 / 2\pi = 0.004 \nu^{1.5}$ to $\Delta \rho [2] / \sqrt{\rho_{\textrm{ref}}}$ \cite{quadspec}.

\begin{figure}
\resizebox{0.50\textwidth}{!}{%
  \includegraphics{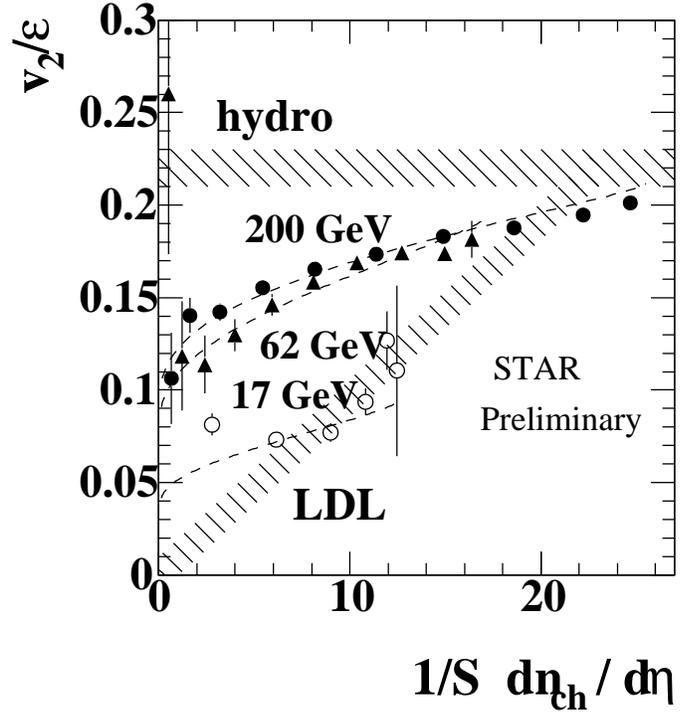}
}
\caption{A hydro-inspired plot of $v_2 / \epsilon$ \emph{vs.} LDL (low-density limit) parameter $1/S \  dn_{ch} / d\eta$ \cite{ldl}.}
\label{fig:energyb}       
\end{figure}

We can contrast this simple description of the data with the hydro-inspired $v_2 / \epsilon$ \emph{vs.} $1/S dn_{ch} / d\eta$ format \cite{ldl} shown in Fig.~\ref{fig:energyb}.  For thermal equilibrium the expectation is the ideal-hydro limit $v_2 / \epsilon \rightarrow \textrm{constant}$. Previous $v_2$ measurements have been interpreted to suggest that central Au-Au collisions at 200 GeV attain the ideal hydro limit (full thermalization over some substantial volume) \cite{volconf}.  The present analysis is \emph{inconsistent with such expectations}. We find a universal linear trend on $n_{\textrm{bin}}$ and $\sqrt{s_{NN}}$ for all data as in Fig.~\ref{fig:energya}.

\section{Minijets}
\label{mjconc}

\begin{figure}
\resizebox{0.50\textwidth}{!}{%
  \includegraphics{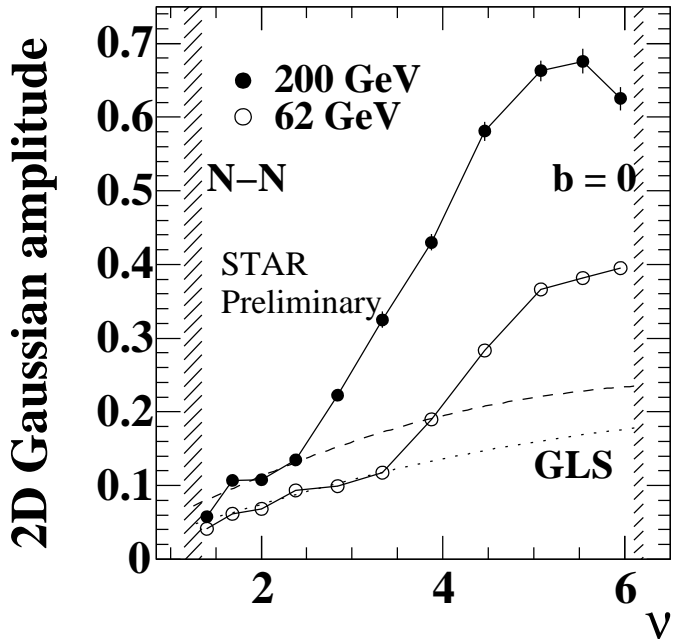}
}
\caption{Minijet peak amplitudes \emph{vs.} mean participant path length $\nu$ for 200 GeV Au-Au collisions (solid dots) and 62 GeV Au-Au collisions (open dots).  The dashed and dotted lines represent binary collision scaling estimates.}
\label{fig:mjtrana}       
\end{figure}

\begin{figure}
\resizebox{0.50\textwidth}{!}{%
  \includegraphics{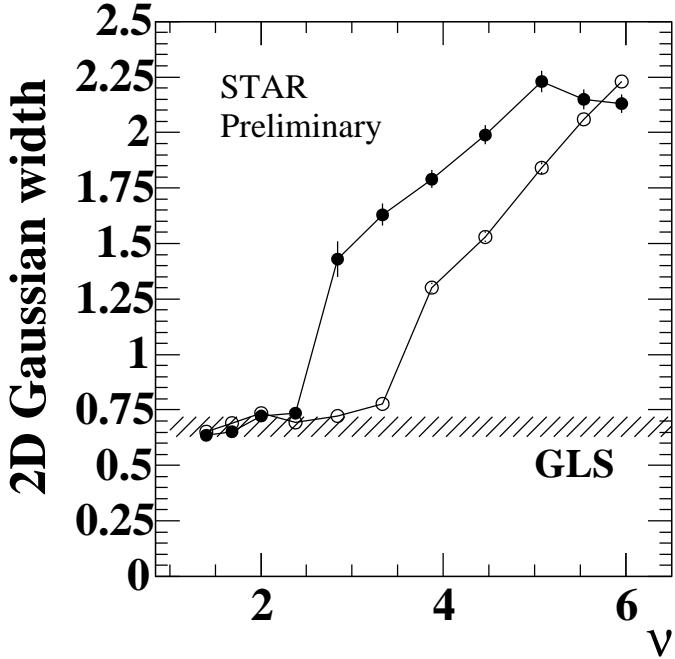}
}
\caption{Minijet peak widths in $\eta_\Delta$ \emph{vs.} $\nu$.}
\label{fig:mjtranb}       
\end{figure}

We have also fit the same-side minijet peak in the same 62 and 200 GeV Au-Au autocorrelations.  The simple behavior of the quadrupole component can be contrasted with the minijet peak, which exhibits a remarkable transition in excess of binary collision scaling  around $\nu = 2.5$ at 200 GeV and $\nu = 3.4$ at 62 GeV \cite{mikeQM} as seen in Fig.~\ref{fig:mjtrana} and Fig.~\ref{fig:mjtranb}.  This deviation is a strong indication of medium effects.  In more peripheral cases the behavior approximates a simple superposition of nucleon-nucleon collisions.  In the quadrupole component no such transition is observed.  We have already seen in Fig.~\ref{fig:centralitya} that there is a smooth evolution according to the collision geometry from the most peripheral to the most central collisions, and this evolution is described by the parametrization in Eq.~(\ref{eqn:scaling}).

From our fits we conclude that the $m=2$ Fourier component of the Gaussian minijet peak is the primary source of nonflow effects in $v_2 \{ 1D \}$ and similar measures.  The trend in Fig.~\ref{fig:mjtrana} also implies that the magnitude of the nonflow contribution is strongly centrality-dependent.

\section{Conclusion}
\label{concl}

In conclusion, azimuth quadrupole moments \emph{vs.} centrality for Au-Au collisions were measured at 62 and 200 GeV separately from minijet contributions (nonflow) via fits to 2D angular autocorrelations.  These fits use the shape of the correlation structures to quantify pair-wise relative pseudorapidity dependent nonflow contributions to two-particle $v_2$.

Using the optical Glauber model of eccentricity we find that the trend $\Delta \rho [2] / \sqrt{\rho_{\textrm{ref}}} \propto R(\sqrt{s_{NN}}) n_{bin} (b) \epsilon_{\textrm{opt}}^2 (b)$ with $R(\sqrt{s_{NN}}) \propto \log (\sqrt{s_{NN}}/13 \textrm{ GeV})$ describes the data for all energies and centralities.  All $p_t$-integrated Au-Au azimuth quadrupole data from 13 to 200 GeV are represented by two initial-state parameters.  This description does not rely on the standard hydrodynamical expectation of an equation-of-state, which we see no evidence for.

We thank the RHIC Operations Group and RCF at BNL, and the NERSC Center at LBNL for their support.  This work was supported in part by the Offices of NP and HEP within the U.S. DOE Office of Science; the U.S. NSF; the BMBF of Germany; CNRS/IN2P3, RA, RPL, and EMN of France; EPSRC of the United Kingdom; FAPESP of Brazil; the Russian Ministry of Sci. and Tech.; the Ministry of Education and the NNSFC of China; IRP and GA of the Czech Republic, FOM of the Netherlands, DAE, DST, and CSIR of the Government of India; Swiss NSF; the Polish State Committee for Scientific Research; Slovak Research and Development Agency, and the Korea Sci. \& Eng. Foundation.


\begin{thebibliography}{}
%
%
\bibitem{classicv2}
J. Adams \emph{et al.} (STAR Collaboration), Nucl. Phys. A \textbf{757}, 102 (2005).
\bibitem{fluctinv}
T. A. Trainor, R. J. Porter, D. J. Prindle, J. Phys. G \textbf{31}, 809 (2005).
\bibitem{poskvol}
A. M. Poskanzer and S.A. Voloshin, Phys. Rev. C \textbf{58}, 1671 (1998).
\bibitem{conf2005}
R. J. Porter and T. A. Trainor (STAR Collaboration), J. Phys. Conf. Series \textbf{27}, 98 (2005).
\bibitem{mj1}
J. Adams \emph{et al.} (STAR Collaboration), Phys. Rev. C \textbf{73}, 064907 (2006).
\bibitem{mj2}
J. Adams \emph{et al.} (STAR Collaboration), J. Phys. G \textbf{32}, L37 (2006).
\bibitem{mj3}
J. Adams \emph{et al.} (STAR Collaboration), J. Phys. G \textbf{33}, 451 (2007).
\bibitem{frag}
T. A. Trainor and D. T. Kettler, Phys. Rev. D \textbf{74}, 034012 (2006).
\bibitem{azstruct}
T. A. Trainor and D. T. Kettler, Int. J. Mod. Phys. E \textbf{17}, 1219 (2008).
\bibitem{quadspec}
T. A. Trainor, Mod. Phys. Lett. A \textbf{23}, 569 (2008).
\bibitem{clusters}
N. Borghini, P.M. Dinh and J.Y. Ollitraut, Phys. Rev. C \textbf{64}, 054901 (2001).
\bibitem{centralities}
T. A. Trainor and D.J. Prindle, hep-ph/0411217
\bibitem{tpcs}
K. H. Ackermann \emph{et al.}, Nucl. Instrum. Meth. A \textbf{499}, 624 (2003).
\bibitem{na49}
A. M. Poskanzer \emph{et al.} (NA49 Collaboration), Nucl. Phys. A \textbf{661}, 341 (1999).
\bibitem{v24}
C. Adler \emph{et al.} (STAR Collaboration), Phys. Rev. C \textbf{66}, 034904 (2002).
\bibitem{v2fluct}
P. Sorensen (STAR Collaboration), nucl-ex/0612021; B. Alver \textbf{et al.} (PHOBOS Collaboration), nucl-ex/0702036.
\bibitem{ecc}
H. Sorge, Phys. Rev. Lett. \textbf{82}, 2048 (1999).
\bibitem{part}
R. S. Bhalerao and J. Y. Ollitrault, Phys. Lett. B \textbf{641}, 260 (2006).
\bibitem{phobscale}
B. Alver (PHOBOS Collaboration), Phys. Rev. Lett. \textbf{98}, 242302 (2007).
\bibitem{jaccoop}
P. Jacobs and G. Cooper, nucl-ex/0008015v1
\bibitem{ldl}
S. A. Voloshin and A. M. Poskanzer, Phys. Lett. B \textbf{474}, 27 (2000).
\bibitem{volconf}
S. A. Voloshin (STAR Collaboration), AIP Conf. Proc. \textbf{870}, 691 (2006).
\bibitem{mikeQM}
Michael Daugherity (STAR Collaboration), nucl-ex/0806.2121
\end{thebibliography}
\end{document}